 \documentstyle[12pt]{article}
 \newcommand{\be}{\begin{equation}}
 \newcommand{\ee}{\end{equation}}
 \newcommand{\ba}{\begin{eqnarray}}
 \newcommand{\ea}{\end{eqnarray}}
 \makeatletter
 \@addtoreset{equation}{section}
 \makeatother
 
 \def\FP{Fad\-deev-\-Po\-pov}
 \def\Haj{H\'aj\'\i\v cek}
 \def\Schr{Sch\-r\"o\-din\-ger}
 \def\Sh{Shan\-mu\-ga\-dha\-san}
 
 \def\entry#1#2{\vbox{\hbox to 112truept{\hrulefill}\break%
                \hbox{\vrule\vbox to 28truept{%
                \vfill%
                \hbox to 112truept{\hfill\quad\small #1\quad\hfill}\break%
                \vfill%
                \hbox to 112truept{\hfill\quad\small #2\quad\hfill}%
                \break\vfill%
                \hbox to 112truept{\hrulefill}}\vrule}}}%

 \def\arrwv#1#2{\vbox to 42truept{\vfill%
               \hbox to 112truept{\put(56,0){\line(0,-1){10}}\hfill}\break%
                \vfill%
                \hbox to 112truept{\hfill\small #1\hfill}\break%
                \vfill%
                \hbox to 112truept{\hfill\small #2\hfill}\break%
                \hbox to 112truept{%
                \put(56,0){\vector(0,-1){10}}\hfill}\break
                \vfill}}

 \def\arrwh#1#2{\vbox to 28truept{\vfill
                \hbox to 92truept{\small\hfill #1\hfill}\break
                \hbox to 92truept{\rightarrowfill}\break
                \hbox to 92truept{\small\hfill #2\hfill}\break\vfill}}
 \def\arrwhdd#1#2{\vbox to 28truept{\vfill
                \hbox to 92truept{\small\hfill #1\hfill}\break
                \hbox to 92truept{\leftarrowfill\hskip -1truemm 
                \rightarrowfill}\break
                \hbox to 92truept{\small\hfill #2\hfill}\break\vfill}}

 \def\vf{\varphi}
 \def\d{\partial}
 \def\arcth{{\rm arctanh\,}}
 \def\arcsinh{{\rm arcsinh\,}}
 \def\de{\Delta_{FP}}
 
 \def\PR#1#2#3{{\it Phys.\ Rev.} {\bf #1}, #2 (#3)}
 \def\PRD#1#2#3{{\it Phys.\ Rev.} {\bf D#1}, #2 (#3)}

 \def\PLB#1#2#3{{\it Phys.\ Lett.} {\bf B#1}, #2 (#3)}
 \def\PRSLA#1#2#3{{\it Proc.\ Royal Soc.\ of London} {\bf A#1}, #2 (#3)}

 \def\JMP#1#2#3{{\it J.\ Math.\ Phys.} {\bf #1}, #2 (#3)}

 \def\IJMPA#1#2#3{{\it Int.\ J.\ Mod.\ Phys.} {\bf A#1}, #2 (#3)}

 \def\CQG#1#2#3{{\it Class.\ Quantum Grav.} {\bf #1}, #2 (#3)}
 \def\GRG#1#2#3{{\it Gen.\ Rel. Grav.} {\bf #1}, #2 (#3)}

\begin{document}
\title{Approximate Canonical Quantization for Cosmological Models}
\author{Marco Cavagli\`a\thanks{E-mail: cavaglia@aei-potsdam.mpg.de; web page: 
http://www.aei-potsdam.mpg.de/\~{}cavaglia}\\
\it
Max-Planck-Institut f\"ur Gravitationsphysik\\ \it
Albert-Einstein-Institut\\ \it
Schlaatzweg 1\\ \it
D-14473 Potsdam, Germany}
\date{\today }
\maketitle

\begin{abstract}

In cosmology minisuperspace models are described by nonlinear
time-re\-pa\-ra\-me\-tri\-za\-tion invariant systems with a finite number of
degrees of freedom. Often these models are not explicitly integrable and
cannot be quantized exactly. Having this in mind, we present a scheme for the
(approximate) quantization of perturbed, nonintegrable, 
time-reparametrization
invariant systems that uses (approximate) gauge invariant quantities. We apply
the scheme to a couple of simple quantum cosmological models.

\medskip\noindent
Pac(s) numbers: 03.20.+i, 04.60.Ds, 98.80.Hw, 04.60.Kz.\\
Keyword(s): Canonical Quantization, Quantum Cosmology, Minisuperspace
Models.

\end{abstract}


\section{Introduction\label{intr}}
Historically, the first attempt in quantizing gravity moved from the
canonical formalism. The origin of quantum gravity dates back to
the pioneeristic works of Bergmann and Goldberg \cite{bergmann} and Dirac
\cite{dirac58} who first investigated in the late 50's the canonical
analysis of classical general relativity and applied to the gravitational
field the theory of quantization of constrained systems \cite{dirac64}. 
This line of research led to the Hamiltonian formulation of the Einstein
theory by Arnowitt, Deser, and Misner \cite{ADM} and to the application of
Dirac quantization to the ADM formalism by Wheeler \cite{wheeler68} and
DeWitt \cite{dewitt67} (Wheeler-DeWitt equation).

In spite of an effort lasting more than 40 years, the conceptual problems
arising in the quantum canonical formulation of gravity are hale and hearty 
and
a large amount of time is presently devoted to the discussion of ``structural
issues in quantum gravity'' \cite{isham95}. Let us just mention the so-called
``problem of time'' \cite{kuch92,isham92}: since the first-class Hamiltonian
for general relativity is identically zero, due to general covariance, the
evolution of the system is ``hidden'' in the constraints.

Minisuperspace models are the natural arena to investigate these structural
issues. By retaining only a finite number of degrees of freedom quantum field
theory reduces to quantum mechanics and typical problems due to the field
nature of the system (for instance, anomalies -- see Ref.\ \cite{BJL})
disappear.  However, other conceptual problems, as the mentioned ``problem of
time'', survive to the reduction to a finite number of degrees of freedom
\cite{CDFm,CDs}. The standard techniques of quantization of constrained gauge
systems cannot be naively applied, for instance, even to the simple case of a
Friedmann-Robertson-Walker (FRW) universe minimally coupled to a massive 
scalar
field:  for nonintegrable systems the absence of a global time parameter
forbids the removal of the gauge degree of freedom and a sensible quantum
theory cannot be obtained. So some approximation must be introduced -- at some
stage -- in the quantization procedure.  A possible approach is given by the
semiclassical approximation.  In this framework a (approximate) \Schr\ 
equation
for the system can be derived \cite{vil89}.  However, the definition of time 
in
the \Schr\ equation -- and the \Schr\ equation itself -- depend on the
particular wave function considered. As a consequence, it is not obvious how 
to 
construct a Hilbert space.

In this paper we present a method to obtain an approximate global time
definition and approximate gauge invariant quantities for finite-dimensional,
nonintegrable, perturbed systems, i.e.\ systems whose nonintegrable sector
can be written as a small perturbation. We develop a scheme that allows the
explicit construction of these quantities to any order in the perturbation
parameter $\lambda$. Then the approximate system to the order $O(\lambda^n)$
can be quantized using the standard operator approach. Finally, we illustrate
the method discussing a couple of quantum cosmological models.
\section{Quantization of Integrable Finite-Dimensional Time-Reparametrization 
Invariant Systems\label{QIS}} 
As a warming up exercise let us discuss finite-dimensional systems that are
explicitly integrable. Due to time-reparametrization invariance their action
reads
\be
S=\int_{t_1}^{t_2} dt\,\left[\dot q_\mu p_\mu-u(t)H(q,p)\right]\,,
\label{action-int} 
\ee 
where $\mu=0,1...N-1$ and $u(t)$ is a nondynamical variable that imposes the
constraint $H=0$.  It is straightforward to see that the action
(\ref{action-int}) remains invariant under the time redefinition $t\to\bar
t=f(t)$ \cite{HT}.  This property implies that the system is invariant under 
the
(gauge) transformation generated by $H$
\ba 
&&\delta q_\mu=\epsilon
{{\d H} \over {\d p_\mu}}=\epsilon\bigl[q_\mu,H \bigr]_P\,,\nonumber\\ 
&&\delta
p_\mu=-\epsilon {{\d H} \over {\d q_\mu}}=\epsilon\bigl[ p_\mu,H
\bigr]_P\,,\label{gauge-tr}\\ 
&&\delta u = {{d \epsilon} \over {dt}}\,,\nonumber
\ea 
where $\epsilon(t_1)=\epsilon(t_2)=0$. Note that the gauge transformation 
(\ref{gauge-tr}) has the same content of the equations of motion. Since the 
system is integrable Eqs.\ (\ref{gauge-tr}) can be explicitly 
solved.

The standard operator approaches to the quantization of the system
(\ref{action-int}) are the Dirac method and the reduced method. In both
approaches the redundant gauge degree of freedom is eliminated via the
imposition of an extra condition $F(q,p;t)=0$ (gauge identity) \cite{rov}: 
in the Dirac method the constraint $H=0$ is promoted to a quantum operator
and then the gauge degree of freedom is eliminated by the gauge fixing;
conversely, the reduced method leads first to a classical reduced gauge-fixed
phase space where quantization can be carried out as usual
(Schr\"odinger equation) and wave functions have the customary
interpretation. 

Since the gauge fixing condition is given in the form of a local relation, it
may happen that the gauge fixing condition does not globally hold in the 
entire
phase space (Gribov obstruction, see for instance Ref.\ \cite{HT}). So the
quantization cannot be carried out. (Geometrically, the existence of a global
time parameter implies that gauge conditions intersect the gauge orbits on the
constraint surface once and only once.) However, when the system is integrable
it is certainly possible to choose a global gauge fixing.  Indeed, we can
construct a canonical transformation $\{q_\mu,p_\mu\}\to \{Q_\mu,P_\mu\}$ 
where
one of the new canonical coordinates, say $P_0$, coincides with the 
constraint,
its conjugate $Q_0$ transforms linearly under the gauge transformations, and 
the
remaining canonical coordinates are gauge invariant (Shanmugadhasan canonical
variables) \cite{shan}.  The conjugate variable to the Hamiltonian, $Q_0$,
defines a global time parameter and there is no Gribov obstruction.  Let us 
see
this in detail.

Consider Eq.\ (\ref{action-int}) and neglect the constraint $H=0$.  Since the
system is integrable we can write and solve the equation for the Hamilton
characteristic function
\be
H\left(q_\mu,{\d W\over\d q_\mu}\right)=\alpha_0\,,~~~~\to~~~~
W\equiv W(q_\mu,\alpha_\mu)\,.
\label{hamilton-eq}
\ee
where $\alpha_\mu$ are $N$ constants (of motion). The Hamilton characteristic 
function generates a canonical transformation to cyclic coordinates 
$\{Q_\mu,P_\mu\}$ 
\ba
&&Q_0={\d W\over\d\alpha_0}=\tau+\beta_0\,,\label{q0}\\
&&Q_i={\d W\over\d\alpha_i}=\beta_i\,,
~~~~~~~~~~i=1,2,...\label{qi}\\
&&P_\mu=\alpha_\mu\,,\label{pmu}\\
\ea
where
\be
\tau(t)=\int_{t_0}^t u(t')dt'\,.
\ee
The quantities $\{Q_\mu,P_\mu\}$ form a set of \Sh\ variables of the system.
Indeed, since $\beta_\mu$ are constants of motion, $Q_i$ and $P_i$ are 
$2(N-1)$
gauge-invariant quantities, $P_0$ is the Hamiltonian and $Q_0$ transforms
linearly for the gauge transformation generated by $H$.  $\alpha_\mu$ and
$\beta_\mu$ can be written as functions of $q$ and $p$ evaluating Eqs.\
(\ref{q0}-\ref{pmu}) at $\tau=0$ and inverting the relations
\be
\beta_\mu={\d W\over \d\alpha_\mu}\,,~~~~~~p_\mu={\d W\over \d q_\mu}\,.
\label{alpha-beta}
\ee
The canonical transformation to \Sh\ variables finally reads
\be
Q_\mu=\beta_\mu(q,p)\,,~~~~~~P_\mu=\alpha_\mu(q,p)\,.\label{sh-hamilt1}
\ee
The quantities $\{Q_i,P_\mu\}$ form a complete set of gauge-invariant
quantities (observables) that are in a one-to-one correspondence with the
initial conditions. Note that the definition of \Sh\ variables given
by Eq.\ (\ref{sh-hamilt1}) is not unique. For instance, the generator of
the canonical transformation is defined up to an additive generic function
of the new momenta $f(\alpha_\mu)$
\be
P_\mu\to P'_\mu=P_\mu\,,~~~~~
Q_\mu\to Q'_\mu=Q_\mu+{\d f\over \d P_\mu}\,.
\label{multiple-sh}
\ee
Clearly, $\{Q'_\mu, P'_\mu\}$ form a different set of \Sh\
variables.

In the \Sh\ representation (\ref{sh-hamilt1}) the action reads
\be
S=\int dt(\dot Q_0 P_0+\dot Q_i P_i-uP_0)\,.
\label{action-sh}
\ee
The quantity $Q_0$ can be used to fix the gauge because its transformation
properties under the gauge transformation imply that time defined by
this variable covers once and only once the symplectic manifold, i.e.\
time defined by $Q_0$ is a global time (see e.g.\ Ref.\ \cite{haj}). Using the
Shanmugadhasan variables the quantization procedure becomes trivial
and both Dirac and reduced approaches lead to the same Hilbert space. 

It is well-known that classical canonical transformations and canonical
quantization generally do not commute. Graphically (operators are marked 
with the hat symbol)
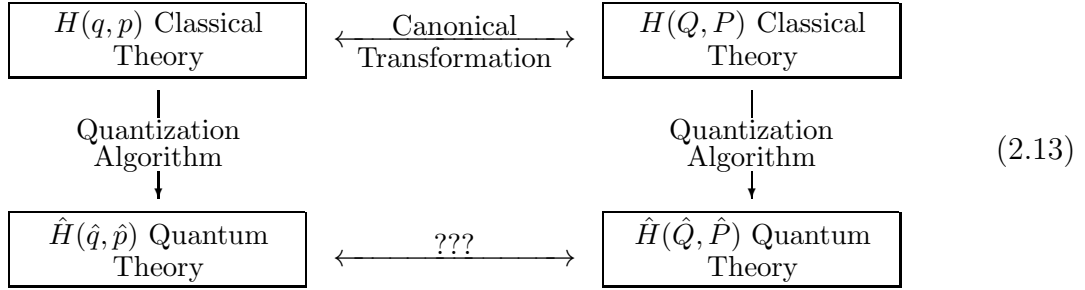
\be
\begin{array}{rcc}
\entry{$H(q,p)$ Classical}{Theory}
&\arrwhdd{Canonical}{Transformation}&\entry{$H(Q,P)$ Classical}{Theory}\\
\arrwv{Quantization}{Algorithm}&&\arrwv{Quantization}{Algorithm}\\
\entry{$\hat H(\hat q,\hat p)$ Quantum}{Theory}&\arrwhdd{???}{}&
\entry{$\hat H(\hat Q,\hat P)$ Quantum}{Theory}
\end{array}
\ee
So the quantization of a classical system described by different -- but
classically equivalent -- sets of canonical coordinates may lead to different
quantum theories. The \Sh\ variables are not immune from this disease:
quantization in the $\{Q,P\}$ representation may be inequivalent to
quantization in the $\{q,p\}$ representation. However, the \Sh\ set of
canonical variables seems to represent a {\it preferred} choice of canonical
coordinates to be used in the quantization procedure. Indeed, using the \Sh\
variables both Dirac and reduced approaches lead to the same Hilbert space.
This property is not true in a generic representation and gives the \Sh\
set of coordinates a preferred status.

In the Dirac approach we start setting the commutation relations
\be
\bigl[\hat Q_\mu,\hat P_\nu\bigr] = i\delta_{\mu\nu}\,.
\label{commutators}
\ee
The quantization of the system can be achieved imposing that the constraint
becomes an operator on wave functions. This enforces the gauge
invariance of the theory.  Once one has imposed the quantum relation $\hat
H\Psi=0$, the gauge must be fixed. This can be done using the
Faddeev-Popov procedure to define the inner product. We have
\be
(\Psi_2,\Psi_1)\,=\,\int d[Q_\mu]\,\Psi_2^*(Q_\mu)\delta (F)
\Delta_{FP} \Psi_1(Q_\mu)\,,
\label{inner}
\ee
where $d[Q_\mu]$ is defined in the unconstrained phase space and represents 
the
measure invariant under the symmetry transformations of the system (rigid and
gauge transformations). $\Delta_{FP}$ is the \FP\ determinant (see e.g.\ Ref.\
\cite{HT}).

Using the measure $d[Q_\mu]=\prod dQ_\mu$ (we neglect possible issues related 
to
the self\--adjoint\-ness nature of the operators) and the operator 
representation
$\hat Q_\mu\to Q_\mu$, $\hat P_\mu\to -i\d_\mu$ the constraint equation reads
$-i\d_0\Psi(Q_i)=0$. So the physical wave functions do not depend on $Q_0$
and the zeroth degree of freedom is pure gauge. A suitable gauge fixing
condition is $F\equiv Q_0-k=0$, where $k$ is a parameter. With this choice the
\FP\ determinant is $\de=1$ and the inner product (\ref{inner}) reads
\be
(\Psi_2,\Psi_1)\,=\,\int \prod_i dQ_i\,\Psi_2^*(Q_j)
\Psi_1(Q_j)\,.
\ee
This completes the quantization of the system. The Hilbert space can also
be obtained reducing first the system to the physical degrees of freedom
by the gauge fixing and then applying the quantization algorithm. In this
case we impose the gauge fixing condition $F\equiv Q_0-t=0$. This
determines the Lagrange multiplier as $u=1$ because of the gauge
equations and of the definition of $Q_0$. Now the gauge fixing can be 
implemented by the (time dependent) canonical transformation
\be
Q_0'=Q_0-t\,,~~~~~P_0'=P_0\,,
\label{sh-can-tr}
\ee
imposing $Q_0'=0$ (gauge fixing) and $P_0'$=0 (constraint).
The effective Hamiltonian on the gauge shell becomes $H_{\rm eff}=P_0=0$ and
the \Schr\ equation reads
\be
i{\d~\over\d t}\psi(t,Q_i)=0\,.
\label{sh-schr}
\ee
Wave functions do not depend on $t\equiv Q_0$ and we recover the
same Hilbert space that we have previously found following the Dirac approach.

We have seen that for integrable systems a maximal set of gauge invariant 
canonical variables can be constructed. Then the quantization is carried out 
using the new variables. Now let us discuss a concrete example. 
Consider the Hamiltonian constraint
\be
H=-{1\over 2}(p_0^2+q_0^2)+{1\over 2}(p_1^2+q_1^2)=0\,.
\label{hamilt-ho}
\ee
(Equation (\ref{hamilt-ho}) describes, for instance, a closed FRW universe
coupled to a homogeneus conformal scalar field. In this case $q_0$ is the 
scale
factor of the metric $a$ and $\chi=q_1/a$ is the conformal scalar field
\cite{CDFm}.)

The Hamilton characteristic function can be chosen (we consider without loss 
of generality $p_0$ and $p_1$ positive defined)
\be
W=\int dq_0\sqrt{2(\alpha_1-\alpha_0)-q_0^2}+\int dq_1
\sqrt{2\alpha_1-q_1^2}\,.
\label{cf-ho}
\ee
Using Eqs.\ (\ref{alpha-beta}) it is straightforward to obtain the \Sh\
variables
\ba
&&Q_0=\arccos{q_0\over\sqrt{p_0^2+q_0^2}}\,,\label{Q0-ho}\\
&&Q_1=-\arccos{q_0\over\sqrt{p_0^2+q_0^2}}-
\arccos{q_1\over\sqrt{p_1^2+q_1^2}}\,,\label{Q1-ho}\\
&&P_0=H(q_i,p_i)\,,\label{P0-ho}\\
&&P_1={1\over 2}(p_1^2+q_1^2)\,.
\label{P1-ho}
\ea
The system can be quantized using $Q_0$ as time parameter. In the original 
variables the time parameter reads
\be
t=\arctan{p_0\over q_0}\,.
\label{time-ho}
\ee
In this case the system is separable and time is defined only
by one of the two degrees of freedom. So we can reduce the system to the
\Sh\ form only in the sector defined by $(q_0,p_0)$. In this
representation the ``hybrid'' canonical variables are
$(Q'_0,P'_0,q_1,p_1)$, where
\ba
&&Q_0'=\arccos{q_0\over\sqrt{p_0^2+q_0^2}}\,,\label{Q0-ho-2}\\
&&P_0'=-{1\over 2}(p_0^2+q_0^2)\,.
\label{P0-ho-2}
\ea
In the hybrid representation the \Schr\ equation reads
\be
i{\d~\over\d 
t}\psi(q_1;t)={1\over 2}\left(-{\d^2~\over\d
q_1^2}+q_1^2\right)\psi(q_1;t)\,,
\label{schr-hyb}
\ee
where we have used the usual representation for the operators $\hat q_1$ and 
$\hat p_1$. The eigenfunctions of the Hamiltonian are
\be
\psi_n(q_1)={1\over\sqrt{\pi^{1\over 2}2^n n!}}H_n(q_1)e^{-q_1^2/2}\,,
\label{sol-hyb}
\ee
where $H_n$ are the Hermite polynomial of order $n$. The eigenfunctions
(\ref{sol-hyb})  form a orthonormal basis in the Hilbert space.
\section{Quantization of Perturbed Nonintegrable Systems\label{QPS}}
Any integrable time-reparametrization invariant system with a finite number of
degrees of freedom can be exactly quantized in the \Sh\ (or hybrid) form. The
canonical coordinate conjugate to the Hamiltonian can be used to define the 
time
parameter for the system and a \Schr\ equation can be written. However, the
quantization procedure illustrated in the previous section relies deeply on
integrability and on the existence of a global time. If the system under
consideration does not admit a global time exact quantization is not possible.
So an approximate method of quantization for these systems is worth exploring.

In this paper we are interested in systems that are described by a Hamiltonian
constraint of the form
\be
H\equiv H_{\rm int}(q_\mu,p_\mu)+\lambda H_{\rm pert}(q_\mu,p_\mu)\,,
\label{pert-hamilt}
\ee
where $\lambda$ is a small dimensionless parameter, $H_{\rm int}$ identifies 
the integrable sector of the theory, and $H_{\rm pert}$ represents the
nonintegrable perturbation.

We have mentioned in the introduction the semiclassical approximation
\cite{vil89}. In this paper we follow a different approach. Our aim is to
obtain approximate gauge invariant quantities for the system, i.e.\ 
approximate
\Sh\ variables up to a certain order $n$ in the parameter $\lambda$. The 
action
is reduced to the form (\ref{action-sh}) plus a term of the order
$O(\lambda^{n+1})$ and the (approximate) system can be quantized along the 
lines
of the previous section. Let us see the procedure in detail.

Since the first term in the Hamiltonian is completely integrable, let us
suppose without loss of generality that the Hamiltonian (\ref{pert-hamilt}) 
is
\be
H\equiv p_0+\lambda H_1(q_\mu,p_\mu)\,.
\label{pert-hamilt2}
\ee
(Is it always possible to reduce Eq.\ (\ref{pert-hamilt}) in the form 
(\ref{pert-hamilt2}) using the canonical transformation described in the
previous section.) The reduction to the \Sh\ form is obtained 
implementing the canonical transformation generated by 
\be
W(q_\mu,P_\mu)=q_\mu P_\mu+\lambda W_1(q_\mu,P_\mu)+\lambda^2 
W_2(q_\mu,P_\mu)+....\,,
\label{generat}
\ee
where $W_1(q_\mu,P_\mu)$, $W_2(q_\mu,P_\mu)$,... are determined by the 
request that the transformed Hamiltonian has the form
\be
H=P_0+O(\lambda^{n+1})\,.
\label{H-transfromed}
\ee
Calculating $Q_\mu$ and $p_\mu$ from Eq.\ (\ref{generat}), expanding in powers 
of $\lambda$, and equating terms of the same order we find
\ba
&&{\d W_1\over\d 
q_0}+H_1(q_\mu,p_\mu)|_{p_\mu=P_\mu}=0\,,\nonumber\\
&&{\d W_2\over\d q_0}+{\d W_1\over\d 
q_\mu}{\d H_1\over\d p_\mu}|_{p_\mu=P_\mu}=0\,,\label{expansion}\\
&&~~~~~~~....\nonumber
\ea
From Eqs.\ (\ref{expansion}) it is straightforward to obtain the
expressions for $W_n$. For instance $W_1$ reads
\be
W_1(q_\mu,P_\mu)=-\int^{q_0}\, dq_0' 
H_1(q_0',q_i,p_\mu)|_{p_\mu=P_\mu}+Z(q_i,P_\mu)\,,
\label{order1}
\ee
where $Z$ is an arbitrary function that we set for simplicity equal to
zero. (Note that all $W_n$ are determined up to additive
arbitrary functions of the new momenta: $W_n\to W_n+Z_n(P_\mu)$ since
Eqs.\ (\ref{expansion}) involve only derivatives of $W_n$ w.r.t.\
$q$.) The first-order canonical transformation is generated by
\be
W(q_\mu,P_\mu)=q_\mu P_\mu-\lambda \int^{q_0} dq_0' \,
H_1(q_0',q_i,p_\mu)|_{p_\mu=P_\mu}+O(\lambda^2)\,.
\label{gen-order1}
\ee
The second term reads
\be
W_2(q_\mu,P_\mu)=\int^{q_0} dq_0' \left\{H_1 {\d H_1\over\d p_0}+ 
\int^{q_0'} dq_0''{\d H_1\over\d q_i}{\d H_1\over \d 
p_i}\right\}_{p_\mu=P_\mu}\,, \label{order2}
\ee
and so on. Implementing the canonical transformation generated by Eq.\ 
(\ref{generat}) to the order $n$ one finds approximate gauge invariant
variables to the order $O(\lambda^n)$. The Poisson brackets read 
\be
[Q_\mu,P_\nu]_P=\delta_{\mu\nu}+O(\lambda^{n+1})\,.
\label{poisson-approx}
\ee
Finally, the canonical variable conjugate to $P_0$
\be
Q_0=q_0+\lambda {\d W_1\over\d P_0}+\lambda^2 {\d^2W_2\over\d P_0^2}+....
\label{time-approx}
\ee
truncated to the order $O(\lambda^n)$ is the approximate time 
parameter of the system. 

We do not expect in general that the series (\ref{time-approx}) are converging
in the entire phase space because a nonintegrable system does not admit a
global time (see e.g.\ Ref.\ \cite{haj}). Nevertheless $Q_0$ defined in Eq.\
(\ref{time-approx}) is a good time parameter in those regions of the phase 
space
where the Poisson bracket of $Q_0$ with the Hamiltonian is positive defined,
i.e.\ when the quantity $O(\lambda^{n+1})$ in Eq.\ (\ref{poisson-approx}) is
less than $1$. This happens usually when the perturbation $H_{\rm pert}$ is
small compared to the integrable part. Conversely, when $H_{\rm pert}\gg 1$
nonperturbative effects are present and the series (\ref{time-approx}) will 
in
general not converge.  Clearly, for $\lambda=0$ the condition is satisfied and
the zero-order term of $Q_0$ in Eq.\ (\ref{time-approx}) is a global time, as
expected because now the system is integrable.

At this point one can quantize the system using the (approximate) \Sh\ 
variables
along the lines of Sect.\ \ref{QIS}. The quantization procedure can be
completed exactly and formally does not depend on the expansion scheme. 
Indeed,
the expansion is chosen such that the new Hamiltonian is $P_0$ and the wave
functions are diagonalized by suitable operators defined by the gauge 
invariant
variables $Q_i$ and $P_i$. The approximation scheme affects only the
definition of the new canonical quantities as functions of the original
variables of the system.  In conclusion, the procedure simply allows to
calculate the finite (approximate) canonical transformation to a set of \Sh\
variables. For nonintegrable, time-reparametrization invariant, perturbed
systems time and gauge invariant quantities are approximate concepts defined
only in suitable regions of the phase space.

In the next section we shall apply the method illustrated above to a simple 
minisuperspace model.
\section{A Flat FRW Universe Coupled to a Scalar Field\label{FRW}}
Let us consider a flat FRW universe coupled to a scalar field $\vf$. In
Planck units ($l_{\rm Planck}=\sqrt{4\pi G/3}$) the action density for
this model reads
\be
S=\int dt\left[-{1\over 2}{a\dot a^2\over 
N}+{1\over 2}{\dot\vf^2 a^3\over N}-Na^3V(\vf)\right]\,,
\label{FRW-action}
\ee
where $a\ge 0$ is the scale factor of the FRW universe, $N$ is the lapse
function, $V(\vf)$ is the potential of the scalar field, and dots
represent derivatives w.r.t. the time $t$. Equation (\ref{FRW-action}) 
can be cast in the Hamiltonian form. Defining $u=N/a^3$ we have
\be
S=\int dt\left\{\dot a p_a+\dot\vf p_\vf-u\left[{1\over 
2}\left(p_\vf^2-a^2p_a^2\right)+a^6V(\vf)\right]\right\}\,. 
\label{FRW-action-hamilt}
\ee
Clearly, the action (\ref{FRW-action-hamilt}) has the form (\ref{action-int});
$u$ is the nondynamical variable.  When the potential is constant the system 
is
separable and integrable; conversely, when the potential for the scalar field 
is
not constant the system is not integrable. However, if the conditions
\be
a^6 V(\vf)\ll a^2p_a^2 \approx p_\vf^2\label{conv-conditions}
\ee
are satisfied, the last term in Eq.\ (\ref{FRW-action-hamilt}) can be
considered as a small perturbation and we can apply the techniques
developed in the previous section. (For instance, a massive minimally
coupled scalar field with potential $V(\vf)=m^2\vf^2/2$ can be described
by a perturbed model when $m\ll 1$.) In this case Eq.\
(\ref{conv-conditions}) defines the region of the minisuperspace where the
perturbative approximation is well-defined. 

Using the canonical transformation
\be
\bar q_0=\ln a\,,~~~~~~\bar p_0=ap_a\,, \label{ct1} 
\ee
the Hamiltonian can be cast in a useful form. We have
\be
H={1\over 2}(\bar p_1^2-\bar p_0^2)+\lambda e^{6\bar q_0}\bar V(\bar
q_1)\,,
\label{hamilt-bar}
\ee
where $\bar q_1=\vf$, $\bar p_1=p_\vf$, and $\bar V(\bar q_1)=V(\bar
q_1)/\lambda$. So the system in the ``barred'' variables is equivalent
to a two-dimensional Klein-Gordon particle with a time-dependent
potential. In these variables the conditions (\ref{conv-conditions}) read
\be
|\lambda| e^{6\bar q_0}\bar V(\bar q_1)\ll \bar p_0^2 \approx \bar
p_1^2\,.
\label{conv-conditions2}
\ee
We shall first discuss the system in the case of constant potential. In
this case the system is integrable and the exact reduction to the \Sh\
form can be found. Then we shall move to the case of a potential of the
form $\bar V(\bar q_1)=\bar q_1^n/n!$ and apply the perturbative
techniques. We shall see that for $n=0$ the perturbative result coincides
with the exact one.
\subsection{Constant Potential: $\bf \bar V(\bar q_1)=1$\label{FRW-CP}}
Let us reduce the system to the \Sh\ form using the method of 
Sect.\ \ref{QIS}. The generator of the canonical transformation is
\be 
W(\bar q_\mu,P_\mu)=\bar q_1\sqrt{P_1+P_0}+{1\over 
3}\sqrt{P_1-P_0}\left(\sqrt{1+{2\lambda\over P_1-P_0} e^{6\bar q_0}}-
\,\arcth{\sqrt{1+{2\lambda\over P_1-P_0} e^{6\bar q_0}}}\right)\,.
\label{cp-gen}
\ee
Using Eq.\ (\ref{cp-gen}) we find
\ba
&&Q_\mu={1\over 2}{\bar q_1\over\bar p_1}\pm {1\over 6\bar 
p_0\sqrt{1-{2\lambda\over\bar p_0^2}e^{6\bar q_0}}}\arcsinh{\sqrt{{\bar 
p_0^2\over2\lambda}e^{-6\bar q_0}-1}}\,,\label{cp-q}\\
&&P_\mu={1\over 2}\left(\bar p_1^2\mp \bar p_0^2\right)\pm\lambda 
e^{6\bar q_0}\,, 
\label{cp-p}
\ea
where different signs refer to ``$0$'' and ``$1$'' variables respectively. In
the next section we shall investigate the case of non-constant potential in 
the
limit of small $\lambda$ and compare the perturbative result to the constant
potential case. So it is convenient to study the behavior of $Q_0$ in the 
limit
of small $\lambda$. It is easy to see that the \Sh\ variables (\ref{cp-q}) are
ill defined for $\lambda\to 0$. This means that this set of variables cannot 
be
used to describe the evolution of the system when the potential is close to
zero. Since the \Sh\ variables are not unique we can redefine them and 
eliminate
the singularity at $\lambda=0$. Recalling that the generator of
the canonical transformation to \Sh\ variables is determined up to a generic
function of the new momenta we define the new generator
\be
W'(\bar q_\mu,P_\mu)=W(\bar q_\mu,P_\mu)-\int dP_0{1\over
12\sqrt{P_1-P_0}}\ln{\left[{2\over\lambda}\left(P_1-P_0\right)\right]}\,.
\label{cp-gen2}
\ee
With this definition the singularity at $\lambda=0$ in the \Sh\ variables 
is removed. $P_\mu$ are clearly unaffected. The conjugate variables read 
now
\be
Q_\mu={1\over 2}{\bar q_1\over\bar p_1}\pm {1\over 6\bar p_0
\sqrt{1-{2\lambda\over\bar p_0^2}e^{-\bar q_0}}}
\left[-3\bar q_0+\ln\left({1+\sqrt{1-{2\lambda\over\bar p_0^2}
e^{6\bar q_0}}\over 2 \sqrt{1-{2\lambda\over\bar p_0^2}
e^{6\bar q_0}}}\right)\right]\,.
\label{cp-q2}
\ee
In the limit of small $\lambda$ we have
\be Q_\mu={1\over 2}\left({\bar q_1\over\bar p_1}
\mp {\bar q_0\over\bar p_0}\right)\mp{\lambda\over 2
\bar p_0^3}e^{6\bar q_0}\left(\bar q_0-{1\over 6}\right)+O(\lambda^2)\,.
\label{cp-q-approx}
\ee
The quantity $Q_0$ in Eqs.\ (\ref{cp-q2}) can be used as time parameter
for the system because $[Q_0,H]_P=1$. The Poisson bracket of the first
order approximation (\ref{cp-q-approx}) with the Hamiltonian is
\be
[Q_\mu, H]_P=\delta_{\mu 0}\mp 9\lambda^2{e^{12\bar q_0}\over
\bar p_0^4}\,.
\label{poiss-approx}
\ee
The region of the phase space where the first order approximation for the 
time parameter holds is then defined by the relation
\be 
1-9\lambda^2{e^{12\bar q_0}\over \bar p_0^4}>0\,,~~~\to~~~ 
|\lambda|e^{6\bar q_0}<{\bar p_0^2\over 3}\,,
\label{reg-approx}
\ee
or, using the original variables,
\be
3|\lambda|a^4<p_a^2\,.
\label{reg-approx2}
\ee
The region of validity of the first order approximation is consistent
with Eqs.\ (\ref{conv-conditions},\ref{conv-conditions2}) therefore
the perturbation techniques can be succesfully implemented. In the next
subsection we discuss the perturbative technique in the case of a
nonconstant potential $\bar V(\bar q_1)=\bar q_1^n/n!$ and recover the
perturbative result of this section in the case $n=0$. 
\subsection{Nonconstant Potential: $\bf \bar V(\bar q_1)=\bar
q_1^n/n!$\label{FRW-NCP}}
Let us apply the formalism of Sect.\ \ref{QPS} to the case $\bar V(\bar
q_1)=
\bar q_1^n/n!$.  Our starting point is the reduction of the Hamiltonian
(\ref{hamilt-bar}) to the form (\ref{pert-hamilt2}). We define the
canonical tranformation to the new variables $(q_\mu,p_\mu)$
\ba
&&\bar q_\mu=(q_1\mp q_0)\sqrt{p_1\mp p_0}\,,\label{can-pert-q}\\
&&\bar p_\mu=\sqrt{p_1\mp p_0}\,.\label{can-pert-p}
\ea
Using Eqs.\ (\ref{can-pert-q},\ref{can-pert-p}) the Hamiltonian is cast into
the form
\be
H=p_0+\lambda e^{6(q_1-q_0)\sqrt{p_1-p_0}}\bar V
\left[(q_1+q_0)\sqrt{p_1+p_0}\right]\,.
\label{frw-ham-pert}
\ee
The first order Hamilton characteristic function can be easily calculated 
from Eq.\ (\ref{gen-order1}). The result is
\be
W(q_\mu,P_\mu)=q_\mu
P_\mu-{\lambda\over n!}{\gamma^n\over(-6)^{n+1}}e^{12\beta q_1}
{\d^n~\over\d\beta^n}\left({e^{-6\beta\xi}\over\beta}\right)+O(\lambda^2)\,,
\label{frw-gen-order1}
\ee
where $\beta=\sqrt{P_1-P_0}$, $\gamma=\sqrt{P_1+P_0}$, and $\xi=q_1+q_0$. 
Starting from Eq.\ (\ref{frw-gen-order1}) the first order gauge invariant
quantities and the first order time parameter can be calculated. We have
\ba
&&Q_0=q_0+{\lambda\over 2(n-1)!}{\gamma^{n-2}\over (-6)^{n+1}}
e^{12\beta q_1}\left[\left({12 q_1\gamma^2\over n\beta}-1\right)
{\d^n~\over\d\beta^n}+{\gamma^2\over n\beta}
{\d^{n+1}~\over\d\beta^{n+1}}\right]
\left({e^{-6\beta\xi}\over\beta}\right)+O(\lambda^2)\,,\nonumber\\
&&\label{frw-q0-approx}\\
&&Q_1=q_1-{\lambda\over 2(n-1)!}{\gamma^{n-2}\over (-6)^{n+1}}
e^{12\beta q_1}\left[\left({12 q_1\gamma^2\over n\beta}+1\right)
{\d^n~\over\d\beta^n}+{\gamma^2\over n\beta}
{\d^{n+1}~\over\d\beta^{n+1}}\right]
\left({e^{-6\beta\xi}\over\beta}\right)+O(\lambda^2)\,,\nonumber\\
&&\label{frw-q1-approx}\\
&&P_0=p_0+{\lambda\over n!}(\gamma\xi)^n 
e^{6\beta(q_1-q_0)}+O(\lambda^2)\,,\label{frw-p0-approx}\\
&&P_1=p_1+{\lambda\over n!}(\gamma\xi)^n
e^{6\beta(q_1-q_0)}-{2\lambda\over n!}{\beta\gamma^n\over (-6)^n}
e^{12\beta q_1}{\d^n~\over\d\beta^n}
\left({e^{-6\beta\xi}\over\beta}\right)+O(\lambda^2)\,,
\label{frw-p1-approx}
\ea
where now $\beta$ and $\gamma$ are evaluated to the order $O(\lambda)$, i.e.\
$\beta=\sqrt{p_1-p_0}$, $\gamma=\sqrt{p_1+p_0}$. In the barred variables 
the above canonical quantities reads
\ba
&&Q_0={1\over 2}\left({\bar q_1\over\bar p_1}-{\bar q_0\over\bar p_0}\right)
+{\lambda\over 2(n-1)!}{\bar p_1^{n-2}\over (-6)^{n+1}}
e^{6\bar p_0\left({\bar q_1\over\bar p_1}+{\bar q_0\over\bar p_0}
\right)}A_n^{(-)} +O(\lambda^2)\,, 
\label{frw-q0-barred}\\
&&Q_1={1\over 2}\left({\bar q_1\over\bar p_1}+{\bar q_0\over\bar p_0}\right)
-{\lambda\over 2(n-1)!}{\bar p_1^{n-2}\over (-6)^{n+1}}
e^{6\bar p_0\left({\bar q_1\over\bar p_1}+{\bar q_0\over\bar p_0}
\right)}A_n^{(+)} +O(\lambda^2)\,, 
\label{frw-q1-barred}\\
&&P_0={1\over 2}(\bar p_1^2-\bar p_0^2)+{\lambda\over n!}\bar q_1^n e^{6\bar 
q_0}+O(\lambda^2)\,,\label{frw-p0-barred}\\
&&P_1={1\over 2}(\bar p_1^2+\bar p_0^2)+{\lambda\over n!}\bar q_1^n e^{6
\bar q_0}-{2\lambda\over n!}{\bar p_0\bar p_1^n\over (-6)^n}
e^{6\bar p_0\left({\bar q_1\over\bar p_1}+{\bar q_0\over\bar p_0}
\right)}B_n+O(\lambda^2)\,,
\label{frw-p1-barred}
\ea
where
\ba
&&A_n^{(\pm)}=\left\{\left[{6\bar p_1^2\over n\bar p_0}\left({\bar 
q_1\over\bar p_1}+ {\bar q_0\over\bar p_0}\right)\pm 1\right]
{\d^n~\over\d\bar p_0^n}+{\bar p_1^2\over n\bar p_0}
{\d^{n+1}~\over\d\bar p_0^{n+1}}\right\}
\left({e^{-6\bar p_0\bar q_1/\bar p_1}\over\bar p_0}\right)\,, 
\label{frw-an}\\
&&B_n={\d^n~\over\d\bar p_0^n}\left({e^{-6\bar p_0\bar q_1/\bar 
p_1}\over\bar p_0}\right)\,. \label{frw-bn}
\ea
After some algebra and using the Poisson brackets
\ba 
&&\left[B_n,\bar p_0^2\right]_P=0\,,
\label{frw-poiss1}\\ 
&&\left[B_n,\bar p_1^2\right]_P=-12\left({-6\bar q_1\over\bar p_1}
\right)^n e^{-6\bar p_0\bar q_1/\bar p_1}\,,
\label{frw-poiss2}\\
&&\left[B_n,{\bar q_0\over \bar p_0}\right]_P=-{1\over\bar p_0}B_{n+1}\,,
\label{frw-poiss3}\\
&&\left[B_n,{\bar q_1\over \bar p_1}\right]_P=0\,,
\label{frw-poiss4}
\ea
it is easy to verify that $[Q_\mu,P_\nu]_P=\delta_{\mu\nu}+O(\lambda^2)$. 
Hence, $Q_0$ defined in Eq.\ (\ref{frw-q0-barred}) is the first order
approximation of the global time parameter of the system and the
quantities defined in Eqs.\ (\ref{frw-q1-barred}-\ref{frw-p1-barred}) form
a complete set of approximate gauge invariant quantities. Setting $n=0$ it
is straightforward to check that Eqs.\
(\ref{frw-q0-barred}-\ref{frw-p1-barred})  reduce to Eqs.\
(\ref{cp-q-approx}) and Eqs.\ (\ref{cp-p}). Therefore the first-order
expansions (\ref{frw-q0-barred}-\ref{frw-p1-barred}) approximate the exact
expression for a constant potential. When the potential is not constant
the result (\ref{frw-q0-barred}-\ref{frw-p1-barred}) is of particular
relevance. In this case the system is not integrable and a global time and
gauge invariant quantities cannot be found in an analitic form. However,
finite expressions like (\ref{frw-q0-barred}-\ref{frw-p1-barred}) can be
found at any order in the $\lambda$ expansion. 

The procedure illustrated above can be easily generalized to the 
case of a closed FRW universe. In this case the Hamiltonian is defined as 
in Eq.\ (\ref{FRW-action-hamilt}) with an extra term $-a^4/2$. Using the 
canonical transformation
\ba
&&\bar q_0={1\over 2}\ln{{a\over p_a+\sqrt{p_a^2+a^2}}}\,,
\label{FRW-closed-q}\\
&&\bar p_0=a\sqrt{p_a^2+a^2}\,,\label{FRW-closed-p}
\ea
the kinetic part becomes equal to the kinetic part of
Eq.\ (\ref{hamilt-bar}) and the (perturbed) potential part reads
\be
H_1=\left({2\bar p_0\over 1+e^{4\bar q_0}}\right)^3
e^{6\bar q_0}\bar V(\bar q_1)\,. \label{FRW-closed-H1}
\ee
The discussion of this case proceeds along the lines of the flat 
FRW case, the only difference being the form of the perturbed Hamiltonian 
in the barred variables.
\section{Conclusions}
In this paper we have illustrated a procedure that allows the approximate
quantization of finite-dimensional, time-reparametrization invariant, non
integrable systems. The method is based on the construction of a set of \Sh\
variables approximated to the order $O(\lambda^n)$ in the perturbation 
parameter
$\lambda$ which identifies the nonintegrable sector of the theory. Using the
new canonical variables the system becomes trivial and quantization can be 
carried out by standard techniques.

The quantization scheme illustrated above has some similarities with the 
method
developed by Barvinsky \cite{Barvinsky} and Barvinsky and Krykhtin \cite{BK}.
In Ref.\ \cite{BK} Barvinsky and Krykhtin prove the equivalence of BFV, Dirac,
and reduced quantization at the one-loop (semiclassical) approximation by the
equivalence of the physical inner product in the two approaches.  In the 
present
paper we have proved the equivalence of Dirac and reduced methods by the same
observation (see below Eq.\ (\ref{sh-schr})).  However, in our case the result
is obtained by considering not a semiclassical expansion, but a perturbative
expansion around the integrable sector of the theory.  Further, the 
equivalence
of Dirac and reduced approaches is proved to an arbitrary order in the
perturbation parameter.  A deeper discussion on the relation between our 
method
and the approach illustrated in Ref.\ \cite{BK} is certainly worth
being investigated.

What are advantages and disadvantages of the quantization scheme presented in
the previous sections?  We have seen that Dirac and reduced quantization
coincide -- to the order $O(\lambda^n)$ -- when (approximate) \Sh\ variables 
are
used. Further, the observation that \Sh\ variables represent a preferred 
choice
of coordinates in the phase space may solve, at least partially, the ambiguity
in the choice of gauge fixing and the problem of inequivalence of quantum
theories obtained starting from different classical sets of canonical
coordinates.  In a sense, \Sh\ variables play the role of the ``Cartesian
coordinates'' in elementary quantum mechanics:  quantization prescriptions 
must
be given in the \Sh\ representation.  One of the main limitations of the 
method
is that the relation between \Sh\ and original canonical variables is 
generally
nonlinear. As a consequence, it is very difficult, if not impossible, to 
write
down quantum operators corresponding to the original fields. However, let us
stress that one should be very careful in assigning a privileged meaning to 
the
original fields. In general relativity no one can identify a priori which
quantities determine the physical properties of the system: it is the nature 
of
the system itself that selects ``the physical'' fields. In the canonical
description of the Schwarzschild black hole, for instance, the relevant field 
is 
not the metric tensor but the mass of the black hole (see Ref.\ \cite{BH}), 
i.e\ 
a gauge invariant, nonlinear, function of the ``original field'' (the metric)
that appears in the Einstein-Hilbert action.

\newpage
\null

\noindent
{\large\bf Acknowledgments}

\noindent
We are grateful to Vittorio de Alfaro, Sasha Filippov, Carlo Rovelli, and Alex 
Vilenkin for interesting discussions and useful suggestions on various 
questions 
connected to the ``problem of time'' in general relativity. This work was 
supported by a Human Capital and Mobility grant of the European Union, 
contract 
number ERBFMRX-CT96-0012.

\end{document}